\documentclass[conference]{IEEEtran}
\usepackage{times,latexsym,amsfonts,amsmath,amssymb,mathrsfs,verbatim,cite}
\usepackage{epsfig,epsf}
\usepackage{graphicx}
\usepackage[dvips]{color}
\usepackage{txfonts}

\def\zZ{{\mathbb Z}}
\def\rR{{\mathbb R}}

\def\pP{{\mathbb P}}

\def\@begintheorem#1#2{\tmpitemindent\itemindent\topsep 0pt\rm\trivlist
    \item[\hskip \labelsep{\indent\it #1\ #2:}]\itemindent\tmpitemindent}
\def\@opargbegintheorem#1#2#3{\tmpitemindent\itemindent\topsep 0pt\rm \trivlist
    \item[\hskip\labelsep{\indent\it #1\ #2\
    \rm(#3):}]\itemindent\tmpitemindent}
\def\@endtheorem{\endtrivlist\unskip}

\newtheorem{theorem}{Theorem}
\newtheorem{definition}{Definition}
\newtheorem{fact}{Fact}

\newtheorem{remark}{Remark}
\renewcommand{\theequation}{\arabic{section}.\arabic{equation}}

\newcommand{\supp}{\operatorname{supp}}
\begin{document}

\title{Infinite-message Interactive Function Computation in Collocated Networks$^{\text{\small 1}}$}


%\author{\authorblockN{Nan Ma and Prakash Ishwar}
%\authorblockA{Department of Electrical and Computer Engineering \\
%Boston University, Boston, MA 02215 \\ {\tt \{nanma, pi\}@bu.edu}
%\\\textbf{Date: \today}
%}}

\author{\IEEEauthorblockN{Nan Ma}
\IEEEauthorblockA{ECE Dept, Boston University\\
Boston, MA 02215 \\ {\tt nanma@bu.edu}}
\and
\IEEEauthorblockN{Prakash Ishwar}
\IEEEauthorblockA{ECE Dept, Boston University\\
 Boston, MA 02215 \\ {\tt pi@bu.edu}}
}

\maketitle
%%%%%%%%%%%%%%%%%%%%%%%%%%%%%%%%%%%%%%%%%%%%%%%%%%%%%%%%%%%%%%%%%%%%%%%%%%%%%%%%
\begin{abstract}
An interactive function computation problem in a collocated network
is studied in a distributed block source coding framework. With the
goal of computing a desired function at the sink, the source nodes
exchange messages through a sequence of error-free broadcasts. The
infinite-message minimum sum-rate is viewed as a functional of the
joint source pmf and is characterized as the least element in a
partially ordered family of functionals having certain
convex-geometric properties. This characterization leads to a family
of lower bounds for the infinite-message minimum sum-rate and a
simple optimality test for any achievable infinite-message sum-rate.
An iterative algorithm for evaluating the infinite-message minimum
sum-rate functional is proposed and is demonstrated through an
example of computing the minimum function of three Bernoulli
sources.
% An interactive function
%computation problem in a collocated network is studied in a
%distributed block source coding framework. With the goal of
%computing a desired function at the sink, the source nodes exchange
%messages through a sequence of error-free broadcasts. Although a
%single-letter characterization of the minimum sum-rate was provided
%in previous work for a finite number of messages, this does not
%leads to a satisfactory characterization of the infinite-message
%limit, which is a new dimension of asymptotic-analysis in
%distributed block source coding. In this paper the infinite-message
%minimum sum-rate is viewed as a functional of the joint source pmf.
%This functional is characterized as the least element in a partially
%ordered family of functionals having certain convex-geometric
%properties. This characterization does not involve taking a limit as
%the number of messages goes to infinity. This characterization leads
%to a family of lower bounds for the infinite-message minimum
%sum-rate and a simple optimality test for any achievable
%infinite-message sum-rate. An iterative algorithm for evaluating the
%infinite-message minimum sum-rate functional is proposed and is
%demonstrated through an example of computing the minimum function of
%three sources.
\end{abstract}
%\vspace{-0.1in}
\section{Introduction}
\addtocounter{footnote}{+1} \footnotetext{This material is based
upon work supported by the US National Science Foundation (NSF)
under award (CAREER) CCF--0546598 and CCF--0915389. Any opinions,
findings, and conclusions or recommendations expressed in this
material are those of the authors and do not necessarily reflect the
views of the NSF. }

In this paper, we study, using a distributed block source coding
framework, an interactive function computation problem in a
collocated network where nodes take turns to broadcast messages over
multiple rounds. Consider a network consisting of $m$ source nodes
and a sink node. Each source node observes a discrete memoryless
stationary source. The sources at different nodes are independent.
The sink does not observe any source and needs to compute a
samplewise function of all the sources. To achieve this objective,
the nodes take turns to broadcast $t$ messages in total. Nodes are
collocated, meaning that every message is recovered at every node
without error. After all the message broadcasts, the sink computes
the samplewise function. The communication is said to be interactive
if $t>m$.

For all finite $t$, a single-letter characterization of the set of
all feasible coding rates (the rate region) and the minimum sum-rate
was provided in \cite{ISIT09} using traditional
information-theoretic techniques. This, however, does not lead to a
satisfactory characterization of the \emph{infinite-message limit}
of the minimum sum-rate as the number of messages $t$ tends to
infinity. The objective of this paper is to provide a ``limit-free''
characterization of the infinite-message minimum sum-rate, i.e., it
does not involve taking a limit as $t\rightarrow \infty$, and also
an iterative algorithm to evaluate it. This result is similar to
that provided in \cite{Allerton09}, where a two-terminal interactive
function computation problem was studied.
 The infinite-message minimum sum-rate is the fundamental
limit of cooperative function computation, where potentially an
infinite number of infinitesimal-rate messages can be used. While
the asymptotics of blocklength, rate, quantizer step-size, and
network size have been explored in the distributed source coding
literature, asymptotics involving an infinite number of messages has
not, to the best of our knowledge, been studied and is not well
understood.

In this paper, we view the infinite-message minimum sum-rate as a
functional of the joint source pmf. The main result is the
characterization this functional as the least element in a partially
ordered family of functionals having certain \emph{convex-geometric}
properties. This characterization does not involve taking a limit as
the number of messages goes to infinity. The proof of this main
result suggests an iterative algorithm for evaluating the
infinite-message minimum sum-rate functional. We demonstrate this
algorithm through an example of computing the minimum function of
three sources.

Related interactive computation problems in various networks have
been studied in
\cite{Kumar2005,GuptaISIT,MassimoAllerton,KumarCDC09} using the
framework of communication complexity \cite{Yao1979,CommComplexity},
where computation is required to be error-free. A function
computation problem in a collocated network is studied in
\cite{Prabha} within a distributed block source coding framework,
under the assumption that conditioned on the desired function, the
observations of source nodes are independent. Multiround
(interactive) function computation in a two-terminal network is
studied in \cite{OrlitskyRoche,ISIT08,Allerton09} within a
distributed block source coding framework.
%There are also some
%papers about computation through noisy channels, but they are
%omitted due to space limitations.
The impact of transmission noise on function computation is
considered in \cite{Gallager88,Srikant,GastparMAC} but without a
block coding rate.

The rest of this paper is organized as follows. In
Sec.~\ref{sec:problem}, we setup the problem and recap previous
results. In Sec.~\ref{sec:characterization} we provide the main
result, a ``limit-free'' characterization of the infinite-message
minimum sum-rate. In Sec.~\ref{sec:itera} we present an iterative
algorithm for evaluating the minimum sum-rate functional and
demonstrate it through an example.
\section{Interactive Computation in Collocated Networks}
\label{sec:problem}

\subsection{Problem formulation}\label{subsec:problemsetup}
Consider a network consisting of $m$ source nodes numbered
$1,\ldots,m$, and an un-numbered sink (node).  Each source node
observes a discrete memoryless stationary source taking values in a
finite alphabet. The sink has no source samples. For each
$j=1,\ldots,m$,
%\footnote{When $a$ and $b$ are integers, $[a,b]$
%denotes an integer interval, which is the set of all consecutive
%integers beginning with $a$ and ending with $b$.}
let $\mathbf X_j:=(X_j(1),\ldots,X_j(n))\in (\mathcal{X}_j)^n$
denote the $n$ source samples which are available at node-$j$. In
this paper, we assume sources are independent, i.e., for
$i=1,\ldots,n$, $(X_1(i),X_2(i),\ldots,X_m(i))$ are iid $p_{X^m}\in
\mathcal{P}_{X^m}$ where $\mathcal{P}_{X^m}:=\left\{ \prod_{j=1}^m
p_{X_j}\right\}$ is the set of all product pmfs on ${\cal X}_1
\times \ldots \times {\cal X}_m$. We adopt this assumption for two
reasons: (1) to isolate the impact of the structure of the desired
function on the efficiency of computation, (2) to obtain an exact
characterization of the optimal efficiency. The general problem
where the sources are dependent across nodes is open. Let $f:
\mathcal{X}_1\times\ldots\times\mathcal{X}_m \rightarrow
\mathcal{Z}$ be the function of interest at the sink and let
$Z(i):=f(X_1(i),\ldots,X_m(i))$. The tuple $\mathbf
Z:=(Z(1),\ldots,Z(n))$, which denotes $n$ samples of the samplewise
function, needs be computed at the sink.

The communication is initiated by node-$k$. The nodes take turns to
broadcast messages in $t$ steps. In the $i$-th step, node-$j$, where
$j=(k+i-1 \mod m)$, \footnote{$j=(k \!\! \mod m)$ means that $j \in
\{1,\ldots,m\}$ and $m$ divides $(j-k)$.} generates a message as a
function of the source samples $\mathbf X_j$ and all the previous
messages and broadcasts it. Nodes are collocated, meaning that every
broadcasted message is recovered without error at every node. After
$t$ message broadcasts, the sink computes the samplewise function
based on all the messages. If $t>m$, the communication is
multi-round and will be called interactive.

%The communication takes place over $r$ rounds. In each round, source
%nodes broadcast messages according to the schedule $1,\ldots,m$.
%Each message depends on the source samples and all the previous
%messages which are available to the broadcasting node. Nodes are
%collocated, meaning that every broadcasted message is recovered
%without error at every node. After $mr$ message broadcasts over $r$
%rounds, the sink computes the samplewise function based on all the
%messages.

\begin{definition}\label{def:code}
A $t$-message distributed block source code for function computation
initiated by node-$k$ in a collocated network with parameters
$(t,n,|{\mathcal M}_1|,\ldots,|{\mathcal M}_t|)$ is the tuple
$(e_1,\ldots,e_t,g)$ consisting of $t$ block encoding functions
$e_1,\ldots,e_t$ and a block decoding functions $g$, of block-length
$n$, where for every $i=1,\ldots,t$, $j=(k+i-1 \!\! \mod m)$,
\begin{equation*}
e_i: \left(\mathcal X_j\right)^{n} \times \bigotimes_{l=1}^{i-1}
{\mathcal M}_{l} \rightarrow {\mathcal M}_i,\ \ \  g:
\bigotimes_{l=1}^{t} {\mathcal M}_l \rightarrow {\mathcal Z}.
\end{equation*}
The output of $e_i$, denoted by $M_i$, is called the $i$-th message.
The output of $g$ is denoted by $\widehat{\mathbf Z}$. For each $i$,
$(1/n) \log_2 |{\mathcal M}_i|$ is called the $i$-th block-coding
rate (in bits per sample).
\end{definition}

\begin{remark}\label{rem:definition}
(i) Each message $M_i$ could be a null message
($|\mathcal{M}_i|=1$). By incorporating null messages, the coding
scheme described above subsumes all orders of messages transfers
from $m$ source nodes, and a $t$-round coding scheme subsumes a
$t'$-round coding scheme if $t'<t$. (ii) Since the information
available to the sink is also available to all source nodes, there
is no advantage in terms of sum-rate to allow the sink to send any
message. (iii) Although the problem studied in \cite{ISIT09} is a
special case with $k=1$ and $t= m r$, where $r\in \zZ^+$ is the
number of rounds, the characterizations for the rate region and the
minimum sum-rate in \cite{ISIT09} naturally extend to the general
problem described above.
\end{remark}

\begin{definition}\label{def:rateregion}
A rate tuple ${\mathbf R} = (R_1, \ldots, R_t)$ is admissible for
$t$-message function computation initiated by node-$k$ if, $\forall
\epsilon
> 0$, $\exists~ \bar n(\epsilon,t)$ such that $\forall n> \bar
n(\epsilon,t)$, there exists a $t$-message distributed block source
code with parameters $(t,n,|{\mathcal M}_1|,\ldots,|{\mathcal
M}_t|)$ satisfying
\begin{equation*}
\forall i =1,\ldots,t, \ \frac{1}{n}\log_2 |{\mathcal M}_i| \leq R_i
+ \epsilon, \ \ \pP(\widehat{\mathbf Z} \neq {\mathbf Z}) \leq
\epsilon.
\end{equation*}
\end{definition}

The set of all admissible rate tuples, denoted by ${\mathcal
R}_t^k$, is called the operational rate region for $t$-message
function computation initiated by node-$k$. The minimum sum-rate
$R_{sum,t}^k$ is given by $\min_{{\mathbf R} \in {\mathcal R}_t^k}
\left(\sum_{i=1}^t R_i\right)$. The focus of this paper is on the
minimum sum-rate rather than the rate region.

\begin{remark} (i) We allow the number of messages $t$ to be equal
to $0$ and abbreviate $R_{sum,0}^k$ to $R_{sum,0}$ because there is
no message transfer and the initial-node is irrelevant. (ii) For
$t<m$, function computation may be infeasible, i.e., $\mathcal
R_t^k$ may be empty. If so, we define $R_{sum,t}^k:=+\infty$. For
special $p_{X^m}$ and $f$, however, computation may be feasible even
with $t<m$; in that case, $R_{sum,t}^k$ would be finite. (iii) For
all $\tau\in \zZ^+$, $R_{sum,t}^k\geq R_{sum,t+\tau}^k\geq 0$ holds,
because the last $\tau$ messages could be null. Hence the limit
$\lim_{t\rightarrow \infty} R_{sum,t}^k =: R_{sum,\infty}^k$ exists
and is finite. (iv) For all $\tau\in \zZ^+$, $R_{sum,t}^k\geq
R_{sum,t+\tau}^{(k-\tau \! \mod m)}$ holds, because the first $\tau$
messages could be null. It follows that $R_{sum,\infty}^k$ is
independent of $k$ and we abbreviate it to $R_{sum,\infty}$. For all
finite $t$, however, we keep the superscript in $R_{sum,t}^k$
because this notation is convenient in the proof of
Theorem~\ref{thm:functioncomp}.
\end{remark}

For all finite $t$, a single-letter characterization of $\mathcal
R_t^k$ and $R_{sum,t}^k$ was provided in Theorem~1 and Corollary~1
of \cite{ISIT09}. This, however, does not directly lead to a
satisfactory characterization of the infinite-message limit
$R_{sum,\infty}$, which is a new dimension for asymptotic-analysis
involving potentially an infinite number of infinitesimal-rate
messages. The main contribution of this paper is a novel
convex-geometric characterization of $R_{sum,\infty}$.

\subsection{Characterization of $R_{sum,t}^k$ for finite $t$}\label{subsec:minsumrate}
\begin{fact}{\it (Characterization of $R^{k}_{sum,t}$
\cite[Corollary~1]{ISIT09})}\label{fact:minsumrate}
\begin{equation}
  R_{sum,t}^k = \min_{p_{U^t|X^m} \in\; \mathcal{P}_t^k(p_{X^m})}
  I(X^m;U^t),\label{eqn:minsumrate}
\end{equation}
where $\mathcal{P}_t^k(p_{X^m})$ is the set of all $p_{U^t|X^m}$
such that (i) $H(f(X^m)|U^t)=0$, (ii) $\forall i\in\{1, \ldots,t\},
j = (k+i-1\mod m), U_i-(U^{i-1},X_j)-(X^{j-1}X_{j+1}^m)$, and (iii)
the cardinalities of the alphabets of the auxiliary random variables
$U^t$ are upper-bounded by functions of $|{\mathcal X}_1|, \ldots,
|{\mathcal X}_m|$ and $t$.
\end{fact}

The Markov chain conditions in Fact~\ref{fact:minsumrate} are
equivalent to the following factorization of $p_{U^t|X^m}$:
\begin{equation}
  p_{U^t|X^m} = p_{U_1|X_k} \cdot
  p_{U_2|X_{(k+1 \!\! \mod m)}U_1}\cdot
p_{U_3|X_{(k+2 \!\! \mod m)} U^2} \ldots. \label{eqn:mc}
\end{equation}
The cardinality bounds in Fact~\ref{fact:minsumrate} which can be
derived using the Carath\'{e}odory theorem are omitted here for
clarity. Although the exact expressions of the cardinality bounds
are unimportant for our discussion, a key property that needs to be
highlighted is that the bound on the alphabet of $U_t$ increases
exponentially with respect to (w.r.t.) $t$. Therefore the dimension
of the optimization problem in \ref{eqn:minsumrate} explodes as $t$
increases.

%\begin{remark}
%When the sources are dependent across nodes, (\ref{eqn:minsumrate})
%is an admissible sum-rate which is not necessarily optimal.
%\end{remark}

Using Fact~\ref{fact:minsumrate}, we could compute $R_{sum,t}^k$ for
a large $t$ to approximate $R_{sum,\infty}$. This is impractical
because (i) the dimension of the optimization problem is large, (ii)
the characterization of $R_{sum,t}^k$ does not inform us how close
$R_{sum,t}^k$ is to $R_{sum,\infty}$. Alternatively, we could
compute $R_{sum,t}^k$ for increasing values of $t$ until
$|R_{sum,t-1}^k-R_{sum,t}^k|$ falls below a threshold. However, the
dimensionality of the optimization problem grows exponentially with
increasing values of $t$ and there is no obvious way to reuse the
computations done for evaluating $R_{sum,t-1}^k$ when evaluating
$R_{sum,t}^k$. Finally, if we need to evaluate $R_{sum,\infty}$ for
a different $p_{X^m}$, we need to repeat the entire process.

In Sec.~\ref{sec:characterization}, we take a new fundamentally
different approach. We first view $R_{sum,\infty}$ as a functional
of $p_{X^m}$ for a fixed $f$. Then we develop a convex-geometric
blocklength-free characterization of the entire functional
$R_{sum,\infty}(p_{X^m})$ which does not involve taking a limit as
$t\rightarrow \infty$. This leads to a simple test for checking if a
given achievable sum-rate functional of $p_{X^m}$ coincides with
$R_{sum,\infty}(p_{X^m})$. It also provides a whole new family of
lower bounds for $R_{sum,\infty}$. In Sec.~\ref{sec:itera}, we use
the new characterization to develop an iterative algorithm for
computing the functional $R_{sum,\infty}(p_{X^m})$ and
$R_{sum,t}^k(p_{X^m})$ (for any finite $t$) in which, crudely
speaking, the complexity of computation in each iteration does not
grow with iteration number, and results from the previous iteration
are reused in the following one. We demonstrate the iterative
algorithm through an example.
\section{Characterization of
$R_{sum,\infty}(p_{X^m})$}\label{sec:characterization}
\subsection{The rate reduction functional $\rho^k_{t}(p_{X^m})$}
\label{subsection:ratered}

If the goal is to {\it losslessly reproduce} the sources, the
minimum sum-rate is equal to $H(X^m)=\sum_{k=1}^m H(X_k)$ because
the sources are independent. The minimum sum-rate for function
computation cannot be larger than that for lossless source
reproduction. The reduction in the minimum sum-rate for function
computation in comparison to source reproduction is given by
\begin{equation}
  \rho^k_{t} :=  H(X^m) - R^{k}_{sum,t} =  \max_{p_{U^t|X^m} \in\;
    \mathcal{P}_t^k(p_{X^m})}H(X^m|U^t).\label{eqn:rho}
\end{equation}
A quantity which plays a key role in the characterization of
$R_{sum,\infty}$ is $\rho_{0}$ -- the ``rate reduction'' for zero
messages (there are no auxiliary random variables in this case). Let
\begin{equation*}
  \mathcal P_{f}:=\left\{p_{X^m}\in \mathcal{P}_{X^m}: H(f(X^m))=0\right\}.
\end{equation*}
Error-free computations can be performed without any message
transfers if, and only if, $p_{X^m}\in\; \mathcal P_{f}$. Thus,
\begin{equation*}
  R_{sum,0}= \left\{
  \begin{array}{cc}
    0, & \mbox{if } p_{X^m}\in\; \mathcal P_{f}, \\
    + \infty, & \mbox{otherwise,}
  \end{array}
  \right.
\end{equation*}
\begin{equation}\label{eqn:rho0}
  \rho_{0}= \left\{
  \begin{array}{cc}
    H(X^m), & \mbox{if } p_{X^m}\in\; \mathcal P_{f},\\
    - \infty, & \mbox{otherwise.}
  \end{array}
  \right.
\end{equation}

\begin{remark}\label{rem:simplexboundary}
If $f(x^m)$ is not constant, for all $p_{X^m}\in \mathcal P_{f}$, we
have $\supp(p_{X^m}) \neq \mathcal X_1 \times \ldots \times \mathcal
X_m$. Such $p_{X^m}$ can only lie on the boundary of
$\mathcal{P}_{X^m}$.
\end{remark}

Evaluating $R^{k}_{sum,t}$ is equivalent to evaluating the rate
reduction $\rho^k_{t}$. It turns out, however, that $\rho_\infty:=
\lim_{t \rightarrow \infty} \rho_t^k=H(X^m)-R_{sum,\infty}$ is
easier to characterize than $R_{sum,\infty}$ (see
Remark~\ref{rem:proofcomments2}). The rate reduction functional is
the key to the characterization.

\subsection{Main result}\label{subsection:mainthm}

Generally speaking, $\rho^k_t$, $\rho_{0}$, and $\rho_{\infty}$
depend on $p_{X^m}$ and $f$. We will fix $f$ and view
$\rho^k_t(p_{X^m})$, $\rho_{0}(p_{X^m})$, and
$\rho_{\infty}(p_{X^m})$ as functionals of $p_{X^m}$ to emphasize
the dependence of $p_{X^m}$. Instead of evaluating
$\rho_{\infty}(p_{X^m})$ for one particular $p_{X^m}$ as it is done
in the numerical evaluation of single-terminal and Wyner-Ziv
rate-distortion functions, our approach is to characterize and
evaluate the functional $\rho_{\infty}(p_{X^m})$ for the entire set
of product distributions $\mathcal P_{X^m}$ rather than for one
particular $p_{X^m}$. To describe the characterization of the
functional $\rho_{\infty}(p_{X^m})$, it is convenient to define the
following family of functionals.

\vspace{1ex}
\begin{definition}{\it (Marginal-\-distributions-\-concave,
    $\rho_0$-\-major- izing family of functionals
    $\mathcal{F}$)}\label{def:F}
  The set of marginal-distributions-concave, $\rho_0$-majorizing family of
  functionals $\mathcal{F}$ is the set of all the
  functionals $\rho: \mathcal P_{X^m} \rightarrow \rR$ satisfying the
  following conditions:
    \begin{enumerate}
    \item $\rho_0$-majorization: $\forall p_{X^m}\in\; \mathcal P_{X^m}$,
      $\rho(p_{X^m})\geq \rho_0(p_{X^m})$.
    \item Concavity w.r.t. marginal distributions: For all $k \in
    \{1,\ldots,m\}$, with $p_{X_j}$ held fixed for all $j\neq k$,
    $\rho\left(\prod_{j=1}^m p_{X_j}\right)$ is a concave function of
      $p_{X_k}$.
    \end{enumerate}
\end{definition}
\vspace{1ex}

\begin{remark}\label{rem:jointfamily}
 Since $\rho_0(p_{X^m})= -\infty$ for all $p_{X^m}\notin \mathcal P_{f}$,
 condition 1) of Definition~\ref{def:F} is trivially satisfied
 for all $p_{X^m}\in\; \mathcal{P}_{X^m} \setminus P_{f}$ (we use
 the convention that $\forall a\in \rR$, $a > -\infty$). Thus the
 statement that $\rho$ majorizes $\rho_0$ on the set $\mathcal
 {P}_{X^m}$ is equivalent to the statement that $\rho$ majorizes
 $H(X^m)$ on the set $\mathcal P_{f}$.
\end{remark}

\begin{remark}\label{rem:jointnonconvex}
 Condition 2) does not imply that $\rho(p_{X^m})$ is concave w.r.t.
 the joint
pmf $p_{X^m}$. In fact, $\mathcal{P}_{X^m}$ is not convex.
\end{remark}

We now state and prove the main result of this paper.

\vspace{1ex}
\begin{theorem}\label{thm:functioncomp}
  (i) $\rho_{\infty} \in\; \mathcal F$. (ii) For all
  $\rho \in\; \mathcal F$, and all $p_{X^m}\in\; \mathcal
  P_{X^m}$, we have $\rho_{\infty}(p_{X^m}) \leq \rho(p_{X^m})$.
\end{theorem}
\vspace{1ex}

The set $\mathcal F$ is partially ordered w.r.t. majorization.
Theorem~\ref{thm:functioncomp} says that $\rho_{\infty}$ is the
least element of $\mathcal F$. Note that there is no parameter $t$
which needs to be sent to infinity in this characterization of
$\rho_{\infty}$.

To prove Theorem~\ref{thm:functioncomp} we will establish a
connection between the $t$-message interactive coding problem and a
$(t-1)$-message subproblem. Intuitively, to construct a $t$-message
interactive code with initial-node $k$ and $p_{X^m}=\prod_{i=1}^m
p_{X_i}$, we need to begin by choosing the first message which
corresponds to choosing the auxiliary random variable $U_1$. Then
for each realization $U_1 = u_1$, constructing the remaining part of
the code becomes a $(t-1)$-message subproblem with initial-node
$k^+:=(k+1 \mod m)$ with the same desired function, but with a
different joint source pmf $p_{X^m|U_1}=\prod_{i=1}^m p'_{X_i}$,
where for all $i\neq k$, $p'_{X_i}=p_{X_i}$ and
$p'_{X_k}=p_{X_k|U_1}$. We can repeat this procedure recursively to
construct a $(t-1)$-message interactive code.  After $t$ steps of
recursion, we will be left with the trivial $0$-message problem.

\begin{proof}
  (i) We need to verify that $\rho_{\infty}$ satisfies the two
  conditions in Definition~\ref{def:F}:

  1) Since $\forall p_{X^m}\in\; \mathcal{P}_{X^m}$,
  $R_{sum,\infty}(p_{X^m})\leq R_{sum,0}(p_{X^m})$, we have
  $\rho_{\infty}(p_{X^m})\geq \rho_0(p_{X^m})$.

  2) For any $k\in \{1,\ldots,m\}$, consider two arbitrary distributions $p_{X_k,0}$ and
  $p_{X_k,1}$, and arbitrary distributions $p_{X_j}$ for all $j\neq
  k$. For $u_1=0,1$, let $p_{X^m,u_1}:=p_{X_k,u_1}\cdot \prod_{j=1,j\neq k}^m
  p_{X_j}$. For $\lambda \in (0,1)$, let $p_{X^m,\lambda}:=
  \lambda p_{X^m,1}+ (1-\lambda ) p_{X^m,0}$. We will show
  that $\rho_{\infty}(p_{X^m,\lambda}) \geq
  \lambda\:\rho_{\infty}(p_{X^m,1})+(1-\lambda)\:\rho_{\infty}(p_{X^m,0})$.
  Let $U_1^*\sim$ Ber$(\lambda)$ and
  $(X^m,U_1^*)\sim p_{X^m,u_1}p_{U_1^*}(u_1)$, which imply $p_{X^m} =
  p_{X^m,\lambda} \in \mathcal{P}_{X^m}$ and
  $p_{X^m|U_1^*}(\cdot|u_1) = p_{X^m,u_1} \in \mathcal{P}_{X^m}$. For all $t\in \zZ^+$ we have,
\begin{eqnarray}
 \rho^k_t(p_{X^m,\lambda}) &=&
      \max_{ p_{U^t|X^m} \in\; \mathcal{P}_t^k(p_{X^m,\lambda}) }
        H(X^m|U^t)\nonumber\\
%\end{eqnarray}
%\begin{eqnarray}
%
  &=&\max_{p_{U_1|X_k}}
       \left\{ \hspace{-10.3ex}
         \max_{ \scriptstyle p_{U_2^t|X^m U_1}: \atop \scriptstyle
       \hspace{10.3ex} p_{U_1|X_k} p_{U_2^t|X^m U_1} \in\;
       \mathcal{P}_t^k(p_{X^m,\lambda}) }
       \hspace{-10.3ex}  H(X^m|U^t)\ \
       \right\} \nonumber \\
  &\stackrel{(a)}{\geq}&\hspace{-1.3ex}\hspace{-10.3ex}
     \max_{ \scriptstyle p_{U_2^t|X^m U_1^*}: \atop \scriptstyle
       \hspace{10.3ex} p_{U_1^*|X_k} p_{U_2^t|X^mU_1^*} \in\;
       \mathcal{P}_t^k(p_{X^m,\lambda}) }
       \hspace{-10.3ex} H(X^m|U_2^t,U_1^*) \nonumber \\
  &\stackrel{(b)}{=}&\hspace{-1.5ex} \lambda \cdot \hspace{-10.3ex}
    \max_{ \scriptstyle \hspace{2ex}
      p_{U_2^t|X^mU_1^*}(\cdot|\cdot,1): \atop \scriptstyle
      \hspace{8.3ex} p_{U_1^*|X_k} p_{U_2^t|X^mU_1^*} \in\;
      \mathcal{P}_t^k(p_{X^m,1}) }
      \hspace{-7ex} H(X^m|U_2^t,U_1^*=1)
       \nonumber \\
  &&\:+ \quad (1-\lambda) \cdot \hspace{-9.5ex}
      \max_{\scriptstyle \hspace{2ex}
    p_{U_2^t|X^m U_1^*}(\cdot|\cdot,0): \atop \scriptstyle
    \hspace{8.3ex} p_{U_1^*|X_k} p_{U_2^t|X^mU_1^*} \in\;
    \mathcal{P}_t^k(p_{X^m,0}) }
  \hspace{-7ex} H(X^m|U_2^t,U_1^*=0) \nonumber
\\
%
%\end{eqnarray}
%\begin{eqnarray}
%
  &\stackrel{(c)}{=}&\lambda\:\rho^{k^+}_{t-1}(p_{X^m,1}) + (1-\lambda)\:
  \rho^{k^+}_{t-1}(p_{X^m,0}).\label{eqn:convexify1}\vspace{-0.1in}
%\\&\stackrel{(b)}{>}& \rho_{\infty}(p_{X,\lambda} \cdot p_{Y|X}).
\end{eqnarray}
 In step (a) we replaced $p_{U_1|X_k}$ with the
particular $p_{U_1^*|X_k}$ defined above. Step (b) follows from the
``law of total conditional entropy'' with the additional
observations that conditioned on $U_1^*= u_1$,
$p_{X^m|U_1^*}(\cdot|u_1) = p_{X^m,u_1}$ and
$H(X^m|U_2^t,U_1^*=u_1)$ only depends on
$p_{U_2^t|X^mU_1^*}(\cdot|\cdot,u_1)$. Step (c) is due to the
observation that for a fixed $p_{U_1^*|X_k}$, conditioned on
$U_1^*=u_1$, $p_{U_1^*|X_k}p_{U_2^t|X^mU_1^*}\in\; \mathcal
P_t^k(p_{X^m,u_1})$ iff $p_{U_2^t|X^mU_1^*}\in\; \mathcal
P_{t-1}^{k^+}(p_{X^m,u_1})$. Now send $t$ to infinity in both the
left and right sides of (\ref{eqn:convexify1}). Since
$\lim_{t\rightarrow \infty} \rho_t^k=\lim_{t\rightarrow \infty}
\rho_t^{k^+}=\rho_{\infty}$, we have
%
%\begin{equation*}
$  \rho_{\infty}(p_{X^m,\lambda})\geq \lambda\:
\rho_{\infty}(p_{X^m,1}
  )+ (1-\lambda)\: \rho_{\infty}(p_{X^m,0}).
$
%\end{equation*}
%
Therefore, $\rho_{\infty}$ satisfies condition 2) in
Definition~\ref{def:F}. Thus, $\rho_{\infty} \in\; \mathcal F$.

(ii) It is sufficient to show that: $\forall \rho \in\; \mathcal F$,
$\forall p_{X^m}\in\; \mathcal P_{X^m}$, $\forall t\in
\zZ^+\bigcup\{0\}$, $\forall k \in \{1,\ldots,m \}$,
$\rho^k_t(p_{X^m})\leq \rho(p_{X^m})$. We prove this by induction on
$t$.  For $t = 0$, the result is true by condition 1) in
Definition~\ref{def:F}. Assume that for an arbitrary $t\in\zZ^+$,
$\rho^k_{t-1}(p_{X^m})\leq \rho(p_{X^m})$ holds. We will show that
$\rho^k_{t}(p_{X^m})\leq \rho(p_{X^m})$ holds.
\vspace{-0.05in}
\begin{eqnarray}
  &&\hspace{-3ex} \rho^k_t(p_{X^m}) \;=
      \max_{ p_{U^t|X^m} \in\; \mathcal{P}_t^k(p_{X^m}) }
      H(X^m|U^t) \nonumber \\
  &=&\max_{ p_{U_1|X_k} }
       \left\{ \hspace{-10ex}
         \max_{ \scriptstyle p_{U_2^t|X^mU_1}: \atop \scriptstyle
       \hspace{10ex} p_{U_1|X_k} p_{U_2^t|X^mU_1} \in\;
       \mathcal{P}_t^k(p_{X^m}) }
           \hspace{-10ex}  H(X^m|U^t)\ \
       \right\} \nonumber
       \\
  &\stackrel{(d)}{=}& \max_{p_{U_1|X_k}}
       \left\{ \hspace{-5ex} \sum_{ \hspace{5ex}u_1\in\;
       \supp(p_{U_1}) } \hspace{-5ex}  p_{U_1}(u_1)
         \left\{ \hspace{-14.5ex}
       \max_{ \scriptstyle \hspace{4ex}
         p_{U_2^t|X^mU_1}(\cdot|\cdot,u_1): \atop
         \scriptstyle \hspace{14.5ex} p_{U_1|X_k} p_{U_2^t|X^mU_1}
         \in\; \mathcal{P}_t^k(p_{X^m|U_1}(\cdot|u_1)) }
         \hspace{-11.5ex} H(X^m|U_2^t,U_1=u_1)
         \right\} \right\} \nonumber\\
  &\stackrel{(e)}{=}& \max_{p_{U_1|X_k}}\left\{ \sum_{u_1\in\;
  \supp(p_{U_1})} p_{U_1}(u_1)\:
  \rho^{k^+}_{t-1}(p_{X^m|U_1}(\cdot|u_1))\right\}\label{eqn:convexify}\\
  &\stackrel{(f)}{\leq}& \max_{p_{U_1|X_k}}\left\{ \sum_{u_1\in\;
  \supp(p_{U_1})} p_{U_1}(u_1)\:
  \rho(p_{X^m|U_1}(\cdot|u_1))\right\}\nonumber
  \end{eqnarray}
\begin{eqnarray}
  &\stackrel{(g)}{=}& \max_{p_{U_1|X_k}}\left\{ \sum_{u_1\in\;
  \supp(p_{U_1})} p_{U_1}(u_1)\:
  \rho(p_{X_k|U_1}(\cdot|u_1)\;
  p_{X^{k-1}X_{k+1}^m})\right\}\nonumber\\
 % \end{eqnarray}
%\begin{eqnarray}
%
  &\stackrel{(h)}{\leq}& \max_{p_{U_1|X_k}}\left\{
  \rho\left(\sum_{u_1\in\; \supp(p_{U_1})}p_{U_1}(u_1)
  p_{X_k|U_1}(\cdot|u_1)\; p_{X^{k-1}X_{k+1}^m}\right)\right\}\nonumber\\
 &=& \rho(p_{X^m}).\nonumber
\end{eqnarray}
The reasoning for steps (d) and (e) are similar to those for steps
(b) and (c) respectively in the proof of part (i) (see equation
array (\ref{eqn:convexify1})). In step (e) we need to use the fact
that $p_{X^m|U_1}(\cdot|u_1) \in \mathcal{P}_{X^m}$, which is due to
(\ref{eqn:mc}) and the assumption that $p_{X^m}\in
\mathcal{P}_{X^m}$. Step (f) is due to the induction hypothesis
$\rho^k_{t-1}(p_{X^m})\leq \rho(p_{X^m})$ for all $k$. Step (g) is
due to the Markov chain $U_1 - X_k - (X^{k-1} X_{k+1}^m)$ and
because $X_k$ and $(X^{k-1} X_{k+1}^m)$ are independent. Step (h) is
Jensen's inequality applied to $\rho(p_{X_k}\cdot
p_{X^{k-1}X_{k+1}^m})$ which is concave w.r.t. $p_{X_k}$.
\end{proof}
%

%\begin{remark}\label{rem:proofcomments2}
%It can be verified that $H(X^m)\in\mathcal{F}$. Although both
%$H(X^m)$ and $\rho_{\infty}(p_{X^m})$ are concave w.r.t. marginal
%distributions $p_{X_k}$, it cannot be claimed that
%$R_{sum,\infty}(p_{X^m}) = H(X^m) - \rho_{\infty}(p_{X^m})$ will be
%{\it convex} w.r.t. marginal distributions. For each $t$, $\rho^k_t$
%is the maximum value of $H(X^m|U^t)$, where the auxiliary variables
%$U^t$ appear only as conditioned random variables. This enables us
%to use the ``law of total conditional entropy'' (which corresponds
%to convexification) and arrive at (\ref{eqn:convexify1}) and
%(\ref{eqn:convexify}). Notice, however, that $R_{sum,t}^k$ is the
%minimum value of $I(X^m;U^t)$ over all $U^t$ where $U^t$ are not
%conditioned. Therefore, $R_{sum,t}^k$ cannot be expressed as a
%convex combination of $R_{sum,t-1}^{k^+}$. Due to these reasons,
%although evaluating $\rho_{\infty}$ is equivalent to evaluating
%$R_{sum,\infty}$, the rate reduction functional is the key to the
%characterization.
%\end{remark}

Since every $\rho\in\; \mathcal F$ gives an upper bound for
$\rho_{\infty}$, $(H(X^m)-\rho)$ gives a lower bound for
$R_{sum,\infty}$. This fact provides a method for testing if an
achievable sum-rate functional is optimal. If $R^*(p_{X^m})$ is an
achievable sum-rate functional then $\forall p_{X^m} \in
\mathcal{P}_{X^m}$, $R^*(p_{X^m}) \geq R_{sum,\infty}(p_{X^m})$. If
it can be verified that $\rho^* := (H(X^m) - R^*)\in \mathcal F$,
then by Theorem~\ref{thm:functioncomp}, $R^* = R_{sum,\infty}$.

\section{Iterative algorithm}\label{sec:itera}

Although Theorem~\ref{thm:functioncomp} provides a characterization
of $\rho_{\infty}$ and $R_{sum,\infty}$ that is not obtained by
taking a limit, it does not directly provide an algorithm to
evaluate $R_{sum,\infty}$. To efficiently represent and search for
the least element of $\mathcal F$ is nontrivial because each element
is a functional; not a scalar. The proof of
Theorem~\ref{thm:functioncomp}, however, inspires an iterative
algorithm for evaluating $R_{sum,t}^k$ and $R_{sum,\infty}$.

Equation (\ref{eqn:convexify}) states that $\rho^k_t(p_{X^m})$ is
the maximum value of $\rho\in \rR$ such that $\left(p_{X^m},
\rho\right)$ is a finite convex combination of
$\{(p_{X^m|U_1}(\cdot|u_1)$,
$\rho^{k^+}_{t-1}(p_{X^m|U_1}(\cdot|u_1))\}_{u_1\in\;
\supp(p_{U_1})}$, where $p_{X^m}(\cdot)$ and
$p_{X^m|U_1}(\cdot|u_1)$ have the same marginal distributions
$p_{X_j}$ for all $j\neq k$ and differ only on $p_{X_k}$. Now we fix
the marginal distributions $p_{X_j}$ for all $j\neq k$, and consider
the hypograph of $\rho_{t-1}^{k^+}$ w.r.t. $p_{X_k}$:
$\mbox{hyp}_{p_{X_k}} \rho_{t-1}^{k^+} := \{(p_{X_k},\rho): \rho\leq
\rho_{t-1}^{k^+}(\prod_{i=1}^m p_{X_i})\}$. Due to
(\ref{eqn:convexify}), the convex hull of $\mbox{hyp}_{p_{X_k}}
\rho_{t-1}^{k^+}$ is $\mbox{hyp}_{p_{X_k}} \rho_{t}^{k}$. This
relation enables us to evaluate $\rho_{t}^k$ from
$\rho_{t-1}^{k^+}$: fixing $p_{X_j}$ for all $j\neq k$, $\rho_{t}^k$
is the least concave functional w.r.t. $p_{X_k}$ that majorizes
$\rho_{t-1}^{k^+}$. In the convex optimization literature,
$(-\rho_t^k)$ is called the double Legendre-Fenchel transform or
convex biconjugate of $(-\rho_{t-1}^{k^+})$ \cite{ConvexAnalysis}.
We have the following iterative algorithm.

\textbf{Algorithm to evaluate $R_{sum,t}^k$}
\begin{itemize}

\item \textbf{Initialization:} For all $k=1,\ldots,m$, define $\rho_0^k(p_{X^m}) = \rho_0(p_{X^m})$
by equation (\ref{eqn:rho0})  for all $p_{X^m}$ in $\mathcal
  P_{X^m}=\left\{\prod_{i=1}^m p_{X_i}\right\}$.

\item \textbf{Loop:} For $\tau=1$ through $t$ do the following.\\
For every $k=1,\ldots,m$ do the following.\\
For every set of marginal distributions $\{p_{X_j}\}_{j=1,j\neq
k}^m$ do the following.

\begin{itemize}

  \item Construct $\mbox{hyp}_{p_{X_k}}\rho_{\tau-1}^{k^+}$.

  \item Let $\rho_\tau^k$ be the upper boundary of the convex hull of
    $\mbox{hyp}_{p_{X_k}}\rho_{\tau-1}^{k^+}$.

  \end{itemize}

  \item \textbf{Output:}
    $R_{sum,t}^k(p_{X^m})=H(X^m)-\rho_t^k(p_{X^m})$.

\end{itemize}
%\vspace{1ex}

To make numerical computation feasible, $\mathcal P_{X^m}$ has to be
discretized. Once discretized, however, in each iteration, the
amount of computation is the same and is fixed by the discretization
step-size. Also note that results from each iteration are reused in
the following one. Therefore, for large $t$, the complexity to
compute $R_{sum,t}^k$ grows linearly w.r.t. $t$.

$R_{sum,\infty}$ can also be evaluated to any precision, in
principle, by running this iterative algorithm for $t = 1, 2,
\ldots$, until some stopping criterion is met, e.g., the maximum
difference between $\rho_{t-1}^k$ and $\rho_t^k$ on $\mathcal
P_{X^m}$ falls below some threshold. Developing stopping criteria
with precision guarantees requires some knowledge of the rate of
convergence which is not established in this paper and will be
explored in future work. When the objective is to evaluate
$R_{sum,\infty}(p_{X^m})$ for all pmfs in $\mathcal P_{X^m}$, this
iterative algorithm is much more efficient than using
(\ref{eqn:minsumrate}) to solve for $R_{sum,t}^k$ for each $p_{X^m}$
for $t = 1, 2, \ldots$, an approach which follows the definition of
$R_{sum,\infty}$ literally as the limit of $R_{sum,t}^k$ as
$t\rightarrow \infty$. Our iterative algorithm is based on
Theorem~\ref{thm:functioncomp} which is a characterization of
$R_{sum,\infty}$ {\it without taking a limit involving} $t$.

\noindent \emph{Example: (MIN function)} Take $m=3$ nodes. $X_i\sim
$ Ber$(p_i)$. $f(x^3)=\min_{i=1,2,3} x_i$. The joint pmf $p_{X^3}$
is parameterized by $\mathbf p=(p_1,p_2,p_3)\in [0,1]^3$. It is easy
to see that
\begin{equation*}
  \rho_{0}(\mathbf p)= \left\{
  \begin{array}{cc}
    \sum_{i=1}^3 h_2(p_i), & \mbox{if } \mathbf p\in\; \mathcal P_{f},\\
    - \infty, & \mbox{otherwise,}
  \end{array}
  \right.
\end{equation*}
where $\mathcal P_f=\{\mathbf p:p_1=p_2=p_3=1$ or $p_1 p_2 p_3=0
\}$.

Now let us fix $p_{X_1}$ and $p_{X_2}$ and apply the convex
biconjugate operation on $\rho_{0}$ w.r.t. $p_{X_3}$ to obtain
$\rho_1^3$. Specifically, for every fixed $(p_1,p_2)$, we focus on
$\rho_0$ on the line segment $\{p_1\}\times \{p_2\}\times [0,1]$ and
convexify $\mbox{hyp}_{p_3} \rho_{0}=\{(p_3,\rho):\rho\leq
\rho_0(p_1,p_2,p_3)\}$ to obtain $\mbox{hyp}_{p_3} \rho_{1}^3$. Then
we repeat this procedure but applying the convexification operation
w.r.t. $p_{X_2}$, $p_{X_1}$, etc to obtain $\rho_2^2$, $\rho_3^1$,
etc. In numerical computation, $\mathbf p$ takes values on a
discrete grid where $p_1,p_2,p_3$ are multiples of a finite step
size $\Delta$. The convexification operation involves finding a
convex hull of a finite number of points in a plane.

\begin{figure}[!htb]
\centering
\begin{picture}(7,3.1)
\put(0,0){\includegraphics[scale =
0.5]{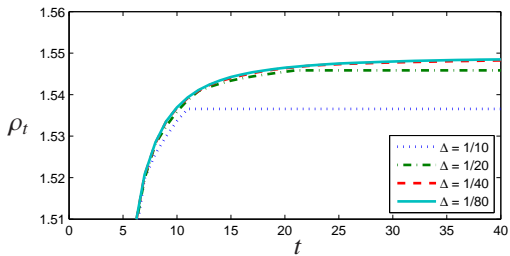}} \put(4,0){\makebox(0,0){$t$}}
\put(0.3,1.6){\makebox(0,0){$\rho_t$}}
\end{picture}
\caption{\sl $\rho_t(1/2,1/2,1/2)$ for different step sizes
$\Delta$\label{fig:stepsize}}
\end{figure}

As we decrease $\Delta$ and increase $t$, $\rho_t$ approximates
$\rho_\infty$. Fig.~\ref{fig:stepsize} shows $\rho_t(1/2,1/2,1/2)$
for different $t$ and $\Delta$. For each $\Delta$, $\rho_t$
converges as $t$ increases. For a small enough $\Delta$ (fine enough
discretization), the limit represents the actual value of
$\rho_\infty(1/2,1/2,1/2)$. Notice that for a small enough $\Delta$,
$\rho_t$ keeps increasing as $t$ grows, which means there is always
an improvement for using more messages.

\begin{figure}[!htb]
\centering
\begin{picture}(7,6)
\put(-0.8,0){\includegraphics[scale =
0.6]{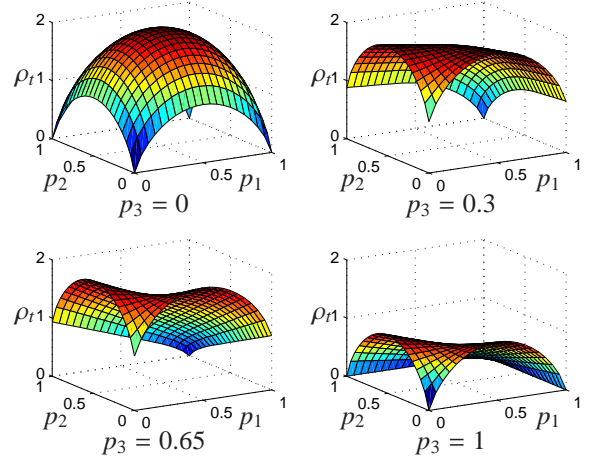}}

\put(0.2,0){\put(2.8,0.6){\makebox(0,0){$p_1$}}
\put(0.2,0.6){\makebox(0,0){$p_2$}}
\put(-0.2,2){\makebox(0,0){$\rho_t$}}
\put(1.5,0.3){\makebox(0,0){$p_3=0.65$}}}

\put(4.15,0){\put(2.8,0.6){\makebox(0,0){$p_1$}}
\put(0.2,0.6){\makebox(0,0){$p_2$}}
\put(-0.2,2){\makebox(0,0){$\rho_t$}}
\put(1.5,0.3){\makebox(0,0){$p_3=1$}}}

\put(0.2,3.16){\put(2.8,0.6){\makebox(0,0){$p_1$}}
\put(0.2,0.6){\makebox(0,0){$p_2$}}
\put(-0.2,2){\makebox(0,0){$\rho_t$}}
\put(1.5,0.3){\makebox(0,0){$p_3=0$}}}

\put(4.15,3.16){\put(2.8,0.6){\makebox(0,0){$p_1$}}
\put(0.2,0.6){\makebox(0,0){$p_2$}}
\put(-0.2,2){\makebox(0,0){$\rho_t$}}
\put(1.5,0.3){\makebox(0,0){$p_3=0.3$}}}

\end{picture}
\caption{\sl Rate reduction function for $t=40$\label{fig:layer}}
\end{figure}

Fig.~\ref{fig:layer} shows the plots of the rate reduction function
with $t=40$ for four values of $p_3$.

%\section{Concluding Remarks}
%
%We studied function computation in collocated networks using a
%distributed block source coding framework. We showed that in
%computing symmetric functions of binary sources, the sink will
%inevitably obtain certain additional information which is not part
%of the problem requirement. Leveraging this conceptual understanding
%we developed bounds for the minimum sum-rate and showed that they
%can be orderwise better than the cut-set bounds.
%%by orders of magnitude
%Directions for future work include characterizing the scaling law of
%the minimum sum-rate for large source alphabets and general multihop
%networks.

%\vspace{-0.1in}
%%%%%%%%%%%%%%%%%%%%%%%%%%%%%%%%%%%%%%%%%%%%%%%%%%
\appendices
\renewcommand{\theequation}{\thesection.\arabic{equation}}
\setcounter{equation}{0}

%To create space, if needed, we can omit references [6] and [13]-[16]
%and change the wording of paragraph 4 in the Introduction. Another
%option is to keep say [16] and if [16] cites [13], [14], [15] we can
%say "see [16] and references therein".

%%%%%%%%%%%%%%%%%%%%%%%%%%%%%%%%%%%%%%%%%%
\footnotesize

%\addcontentsline{toc}{chapter}{References}
\bibliography{newbibfile}
%%%%%%%%%%%%%%%%%%%%%%%%%%%%%%%%%%%%%%%%%%
\end{document}